\documentclass[11pt,twoside]{article}
\usepackage{./asp2014}

\aspSuppressVolSlug
\resetcounters

\begin{document}

\title{An Hierarchical Approach to Big Data}
\author{M. G. Allen$^1$, P. Fernique$^1$, T. Boch$^1$, D. Durand$^2$, A. Oberto$^1$, B. Merin$^3$, F. Stoehr$^4$, F. Genova$^1$, F-X. Pineau$^1$, J. Salgado$^3$
\affil{$^1$Observatoire astronomique de Strasbourg, Universit\'{e} de Strasbourg, CNRS, UMR 7550, 11 rue de l'Universit\'{e}, F-67000 Strasbourg, France; \email{mark.allen@astro.unistra.fr}}
\affil{$^2$National Research Council Canada, Canadian Astronomy Data Centre, 5071 W. Saanich Rd., Victoria, BC, Canada; \email{}}
\affil{$^3$ESA, ESAC Science Data Centre, Spain; \email{}}
\affil{$^4$ALMA Regional Centre, Garching, Germany; \email{}}}

\paperauthor{M. G. Allen}{mark.allen@astro.unistra.fr}{orcid.org/0000-0003-2168-0087}{Observatoire astronomique de Strasbourg, UniversitŽ de Strasbourg, CNRS, UMR 7550}{}{Strasbourg}{Observatoire Astronomique de Strasbourg}{67000}{France}
\paperauthor{P. Fernique}{pierre.fernique@astro.unistra.fr}{}{Observatoire astronomique de Strasbourg, UniversitŽ de Strasbourg, CNRS, UMR 7550}{}{Strasbourg}{Observatoire Astronomique de Strasbourg}{67000}{France}
\paperauthor{T. Boch}{thomas.boch@astro.unistra.fr}{}{Observatoire astronomique de Strasbourg, UniversitŽ de Strasbourg, CNRS, UMR 7550}{Observatoire Astronomique de Strasbourg}{Strasbourg}{}{67000}{France}
\paperauthor{D. Durand}{Author3Email@email.edu}{}{National Research Council Canada, Canadian Astronomy Data Centre}{}{Victoria}{BC}{}{Canada}
\paperauthor{A. Oberto}{anais.oberto@astro.unistra.fr}{}{Observatoire astronomique de Strasbourg, UniversitŽ de Strasbourg, CNRS, UMR 7550}{Observatoire Astronomique de Strasbourg}{Strasbourg}{}{67000}{France}
\paperauthor{B. Merin}{bmerin@sciops.esa.int}{}{ESA}{}{}{}{}{Spain}
\paperauthor{J. Salgado}{Jesus.Salgado@sciops.esa.int}{}{ESA}{}{}{}{}{Spain}
\paperauthor{F.~Stoehr}{fstoehr [at] eso [dot] org}{}{ALMA}{ALMA Regional Centre}{Garching}{}{85748}{Germany}
\paperauthor{Francoise Genova}{francoise.genova@astro.unistra.fr}{0000-0002-6318-5028}{Observatoire astronomique de Strasbourg, UniversitŽ de Strasbourg, CNRS, UMR 7550}{Observatoire Astronomique de Strasbourg}{Strasbourg}{N/A}{F-67000}{France}
\paperauthor{F-X. Pineau}{francois-xavier.pineau@astro.unistra.fr}{}{Observatoire astronomique de Strasbourg, UniversitŽ de Strasbourg, CNRS, UMR 7550}{Observatoire Astronomique de Strasbourg}{Strasbourg}{}{67000}{France}

\begin{abstract}
The increasing volumes of astronomical data require practical methods for data exploration, access and visualisation. The Hierarchical Progressive Survey (HiPS) is a HEALPix based scheme that enables a multi-resolution approach to astronomy data from the individual pixels up to the whole sky. We highlight the decisions and approaches that have been taken to make this scheme a practical solution for managing large volumes of heterogeneous data. Early implementors of this system have formed a network of HiPS nodes, with some 250 diverse data sets currently available, with multiple mirror implementations for important data sets. This hierarchical approach can be adapted to expose Big Data in different ways. We describe how the ease of implementation, and local customisation of the Aladin Lite embeddable HiPS visualiser have been keys for promoting collaboration on HiPS.
\end{abstract}

\section{HiPS}
The Hierarchical Progressive Survey (HiPS) is a HEALPix\footnote{\url{http://healpix.sourceforge.net}} \citep{2005ApJ...622..759G}
 based scheme that enables a multi-resolution approach to astronomy data from the individual pixels up to the whole sky \citep{2015A&A...578A.114F}. HiPS was initially developed at the Centre de donn\'{e}es astronomiques de Strasbourg (CDS\footnote{\url{http://cds.unistra.fr}}) to enable the visualisation of all sky imaging survey data in Aladin \citep{2010ASPC..434..163F}. The initial developments have been extended to astronomical source catalogue data, and also to 3-dimensional data cubes. 

The cooperative use of HiPS by a number of data providers has led to a network of distributed HiPS nodes for sharing and mirroring data sets. We attribute the rapidly growing implementation of this system to the convergence of a number of technical and organisation factors that match the needs of the community. In this article we identify the approaches that have been taken to make this scheme a practical solution for managing large volumes of heterogeneous data, and we consider its applicability to large future data sets.

\section{Design and implementation considerations}

The underlying idea of an hierarchical approach to managing and visualising astronomical survey data is that the data can be accessed at the scale and resolution required for a given purpose. With a hierarchical system one can access just the data that is necessary. This has been very successful for visualisation of large surveys in tools such as  Google sky\footnote{\url{http://www.google.com/sky}}, Microsoft World Wide Telescope\footnote{\url{http://www.worldwidetelescope.org}} (WWT),  and Aladin\footnote{\url{http://aladin.unistra.fr}}. The HiPS scheme builds on the concepts for hierarchical visualisation of survey data with the aim of creating a scientifically robust system for hierarchical use of different data types (catalogues and cubes), and to enable easy and practical use of these data in an interoperable way.  

\articlefigure{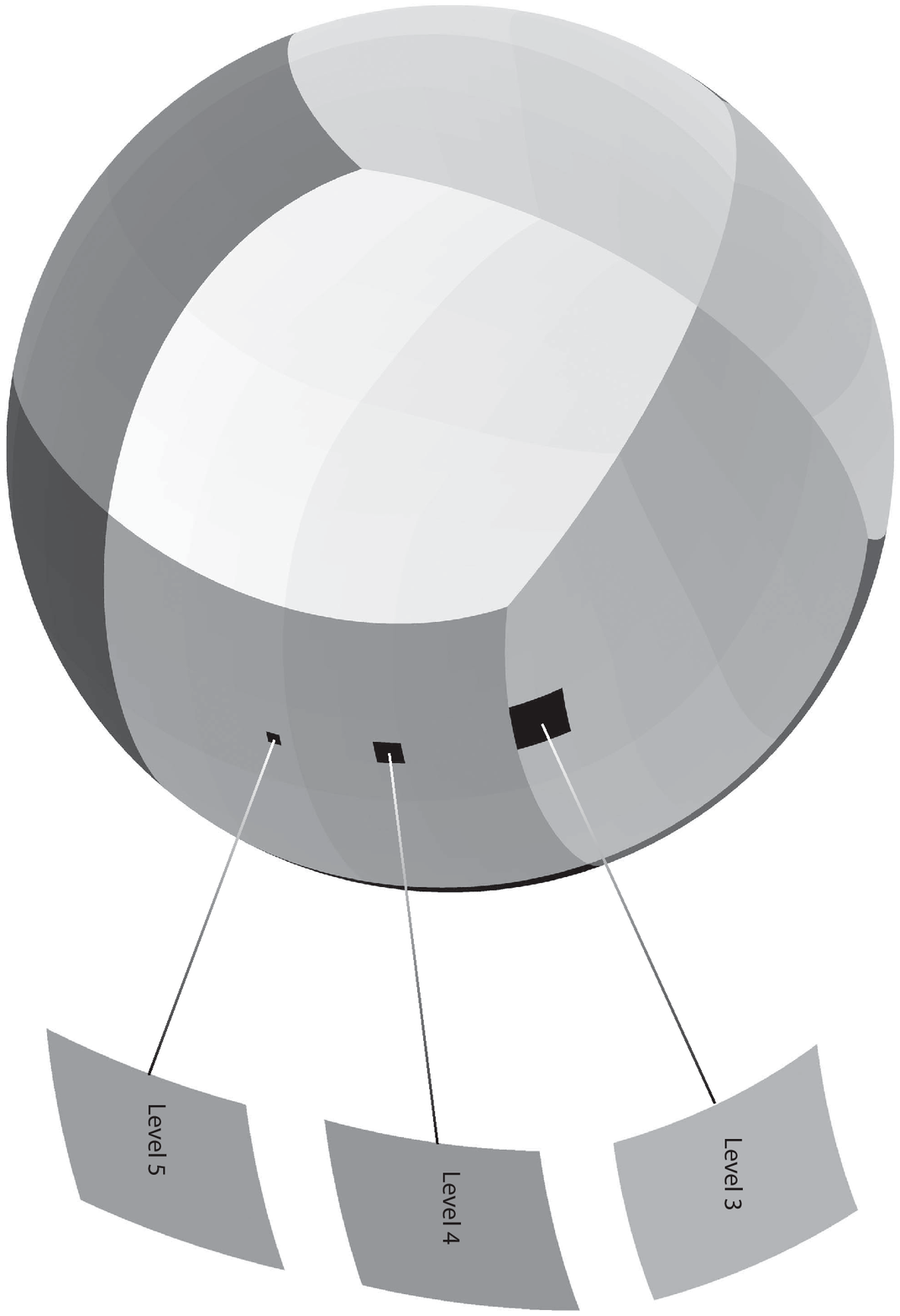}{allen_fig1}{HiPS multi-resolution tile structure based on HEALPix}

HiPS is the result of a long term effort to make data useable and interoperable. The design and implementation considerations that have proven to be successful in the development of HiPS are that: the scientific properties of the original data must be preserved; the system must be simple to use; and that the client tools for use of the system should be independent. 

Preserving the scientific properties of the data has been a key point for going beyond visualisation of the data, and for addressing the requirements for scientific use. The choice of HEALPix as the tessellation for HiPS provides a strong scientific basis because of the existing software and use in the community. Also because of its scientific properties such as equal area pixels and the direct translation of indices into astronomy coordinates, as well as other properties (see -- \citet{2015A&A...580A.132R}). HiPS does of course involve the resampling of original pixel data onto a HEALPix grid, but in order to preserve the original dynamic range of the data values HiPS tiles can be encoded as FITS files (as well as JPEG and PNG). Also, HiPS has an in-built mechanism for the direct referencing of the original data files which contribute to a given HiPS tile, providing important information on the provenance of the data values.

The development of HiPS has involved making choices on the balance of simplicity and complexity in different parts of the system. The structure and encoding of the HiPS data was chosen to be very simple, and to employ standard and pervasive technology. HiPS tiles are stored as simple files in a file system directory structure that is organised into 'HiPS orders' that represent HEALPix maps at different resolutions. This permits direct access to the necessary files for a given region of the sky at the required resolution. The use of simple files and directories (rather than databases) allows HiPS datasets to be published by simply moving the files to a http server. This choice has proved to be extremely useful and well matched to the needs of astronomers and data centre staff as this level of implementation can be performed very easily. An astronomer can create, test and publish a HiPS without need for database expertise, and data centres can test and evaluate use of HiPS with minimal investment or risk. This level of simplicity is possible because file access is sufficiently fast, and because the indexing is implicit in the definition of the HiPS hierarchical directory and file structure. 

The HiPS structure has also proved to be flexible enough to expand its use to other data types based on the concept that a HiPS tile is a generic 'container' (Fig~\ref{allen_fig1}) which may contain any kind of information related to the relevant sky area and resolution level defined by the tile. In this way, the tiles may contain not just images, but data cubes, or catalogues or related provenance links as described in \citet{2015A&A...578A.114F}.

While the HiPS structure itself is very simple, the methods required to generate HiPS data sets are complex. The initial mapping of an image survey into a HEALPix grid can require careful treatment of the mosaicking and background level characteristics of the data. It was a design decision that this complexity be built into the HiPS generation tools. This gives the data providers the flexibility to define and control the way their data is mapped onto a HiPS structure, and provides a way for the data providers to decide the specifics of how the scientific properties of the data are preserved. Significant effort has been put into the development of the {\em hipsgen} tool, in particular to improve the speed of computing the HiPS tiles from the original data. 

There are a number of tools that are available for the use of HiPS. The Aladin desktop java application is a full featured application that has been used as the primary test platform for HiPS. Aladin provides all sky visualisation of HiPS, supporting multiple views with synchronised zoom and pan, and also transparency overlays and interoperability with other online astronomy services. To enable widespread use and access to the tools for generating HiPS, the {\em hipsgen} capabilities are built into Aladin itself, as well as being available as command line programs\footnote{e.g. java -Xmx16000m -jar Aladin.jar -hipsgen in=Fits\_directory}. 

Aladin Lite is a simplified HiPS JavaScript visualiser for web browsers. It is designed to be embedded on a third-party web pages \citep{2014ASPC..485..277B} and it comes with an application program interface (API) that provides the necessary controls to customise it to different needs. Aladin Lite has been particularly important for the implementation of HiPS \citep{P011_adassxxv}. Examples include 'ESA Sky' \citep{P072_adassxxv}, the Spitzer GLIMPSE 360 web page\footnote{\url{http://www.spitzer.caltech.edu/glimpse360/aladin}}  and implementations at CADE\footnote{\url{http://cade.irap.omp.eu}} to preview HEALPix maps of various surveys. Aladin Lite requires only small snipets of code to be embedded, and collaboration with implementors shows that the level of customisation available matches their needs at the right level.

\section{HiPS for Big Data} 

There are currently $\sim$250 HiPS sets available from the CDS and other data providers (see the HiPS directory\footnote{\url{http://aladin.unistra.fr/hips/list}}). This diverse set of HiPS covers a wide range of wavebands, scales and resolution. The recently up-dated HiPS for the HST images available from CADC shows that an entire archive of pointed observations can be combined and used as an image survey. Recent tests with ALMA data cubes show that the calculation and usability of individual 5TB HiPS data sets is feasible.

Since HiPS uses individual files, its scalability for Big Data mainly concerns the issue of scalability and management of file systems that contain many small files, and we expect that generic Big Data solutions for large file systems will be applicable to 'Big HiPS'.  Looking ahead, this scalability provides a way to plan for some aspects of the visualisation and data set exploration challenges that will come with LSST and SKA. A HiPS of a hypothetical LSST data subset covering 18000 square degrees of the sky at 0.3 arc-second resolution, each 3 days for 3 years would lead to a $\sim$5 Petabyte HiPS cube in FITS format, or 256 TB in JPEG format. HiPS access for visualisation of such volumes would be feasible today.

\acknowledgements
MA aknowledges support from the Astronomy ESFRI and Research Infrastructure Cluster -- ASTERICS project, funded by the European Commission under the Horizon 2020 Programme (GA 653477).

\bibliography{O11-2}  

\end{document}